\documentclass[runningheads]{llncs}
\usepackage[T1]{fontenc}
\usepackage{todonotes}
\usepackage{graphicx}
\newcommand{\Fig}[1]{Figure~\ref{#1}}
\newcommand{\Sec}[1]{Section~\ref{#1}}
\newcommand{\Refer}[1]{Reference~\cite{#1}}

\begin{document}

\title{QMIO: A tightly integrated hybrid HPCQC system}
%
%
\author{Javier Cacheiro\inst{1}\orcidID{0000-0001-5864-283X} \and Álvaro C Sánchez\inst{2}\orcidID{0000-0003-2354-4572} \and
Russell Rundle \inst{3}\orcidID{0000-0001-8292-1329} \and
George B Long \inst{3}\orcidID{0000-0002-1787-9539} \and
Gavin Dold\inst{3}\orcidID{0000-0002-6155-3800} \and
Jamie Friel \inst{3}\orcidID{0000-0002-1328-9961} \and
Andrés Gómez\inst{1}\orcidID{0000-0001-7272-8488}}
\authorrunning{J. Cacheiro et al.}

\institute{Galicia Supercomputing Center (CESGA), Santiago de Compostela, Spain\\\email{info@cesga.es} \and
FSAS International Quantum Center (Fujitsu), Santiago de Compostela, Spain \and
Oxford Quantum Circuits (OQC), United Kingdom\\
\email{}}

\maketitle              

\begin{abstract}
High-Performance Computing (HPC) systems are the most powerful tools that we currently have to solve complex scientific simulations. 
Quantum computing (QC) has the potential to enhance HPC systems by accelerating the execution of specific kernels that can be offloaded to a Quantum Processing Unit (QPU), granting them new capabilities, improving the speed of computation, or reducing energy consumption. 
In this paper, we present QMIO: a state-of-the-art hybrid HPCQC system, which tightly integrates HPC and QC. 
We describe its hardware and software components, the integration middleware, and the lessons learned during the design, implementation, and operation of the system. 

\end{abstract}

\section{Introduction}
High-Performance Computing (HPC) systems are the most powerful tools we have at our disposal for solving complex computational problems \cite{qc_review_2024}.
In the future, Quantum Computing (QC) promises to accelerate certain tasks above the possibilities of what the largest classical HPC systems can do.
In fact, QC has made significant progress in recent years, with several providers granting access to Quantum Computing as a Service (QCaaS) through the cloud, and, more recently, delivering on-premises deployments.

The option to deploy a quantum computer on-premises in supercomputing centers is driven by the new capabilities that it can offer as faster solutions for some specific problems, consuming less energy, producing better quality results, and maybe in the future, solving problems that are unresolvable with classical computers (commonly known as \emph{supremacy}). This deployment will create a new hybrid HPC system with QC acceleration,  what we will refer to as an HPCQC system\footnote{There is not a common convention yet about how to call these systems: HPCQC, HPC-QC, QC/HPC, QHPC, <QC|HPC> among others. In other cases it is called \emph{quantum-centric supercomputing}}.

Due to the potential advantages that an HPCQC system could offer, some studies show that 71\% of HPC centers worldwide plan to deploy on-premises quantum computers by 2026 \cite{iqm_2021} and the possibilities of such an integration for some areas of science \cite{Alexeev2024}. 
In fact, some supercomputer centers have already started deploying on-premises quantum computers and there are initiatives such as EuroHPC that plan to deploy quantum computers across seven supercomputing sites in Europe~\cite{eurohpc_qc_2023,eurohpc_qc_2024}. 
Unfortunately, the integration of HPC and QC systems presents significant challenges. 
Numerous research studies have already explored this area, providing valuable insights and proposing different integration strategies that could be used~\cite{Bartsch2021Q,schulz_hpcqc_review,elsharkawy2023integrationquantumacceleratorshigh,giusto2024dependableclassicalquantumcomputersystems,schulz_hpcqc_2023,ornl_hpcqc_2024,quantum_hpc_middleware_2023,Shehata2024IntegratingQC,qc_for_hpc_2021}.

There is also the limitation imposed by the current state of QC technology that is still in a development stage, known as the Noisy Intermediate-Scale Quantum (NISQ) era. 
Currently quantum computing systems face several issues that limit their capabilities due to short coherence times, readout errors, noise and other technological limitations.
Because of these technological limitations, most quantum algorithms are not yet feasible to run. 
However, there are certain classes of algorithms that can be applied in the current NISQ era, for example the Parametric Quantum Circuit (PQC) techniques as the Variational Quantum Eigensolver (VQE), Quantum Approximate Optimization Algorithm (QAOA), and  Quantum Machine Learning (QML) \cite{quantum_hpc_middleware_2023}, or some versions of the Iterative Quantum Phase Estimation (IQPE)\cite{Cruz2020} or the Quantum-Selected Configuration Interfaces (QCSI)\cite{Kanno2023} methods for diagonalization of Hamiltonians.

In practice, these are all hybrid algorithms, using both a classical and a quantum computing system. In most cases, they require hundreds of thousands of calls from the classical computer to the quantum computer until they converge and reach the result, as well as a big number of repetitions (known as \emph{shots}) to be later processed by the HPC system \cite{Robledo-Moreno2024}.
Due to the high number of quantum kernels needed by PQC algorithms, a tight integration is needed between the classical and the quantum systems, which enables short time-to-solution of each offloaded quantum kernel, to significantly decrease the total time to achieve a solution. In the case of IQPE the result of the execution of one quantum kernel is used to select the next step, requiring a low latency round-trip. Other algorithms need even a tighter integration, as the hybrid quantum-classical neural networks~\cite{Schetakis2022}, which join classical and quantum neural networks in a single model. They need a fast execution of each component both during training and inference steps.

In this work, we explore how this type of tightly integrated quantum-classical system can be implemented in practice.
We introduce QMIO, a hybrid HPCQC system, that integrates a 32-qubit Quantum Processing Unit (QPU), an HPC cluster and a quantum emulator into a tightly coupled HPCQC system.

The paper is organized as follows. In \Sec{sec:hardware} we describe the hardware components of the system, in \Sec{sec:software} we present the software and the integration middleware, in \Sec{sec:operation} we discuss the operation of the system, and finally in \Sec{sec:discussion} we provide a summary of the lessons learned and the future work to improve the system.

\section{HPCQC Hardware}\label{sec:hardware}
\begin{figure}[tb]
    \centering
    \includegraphics[width=0.85\linewidth]{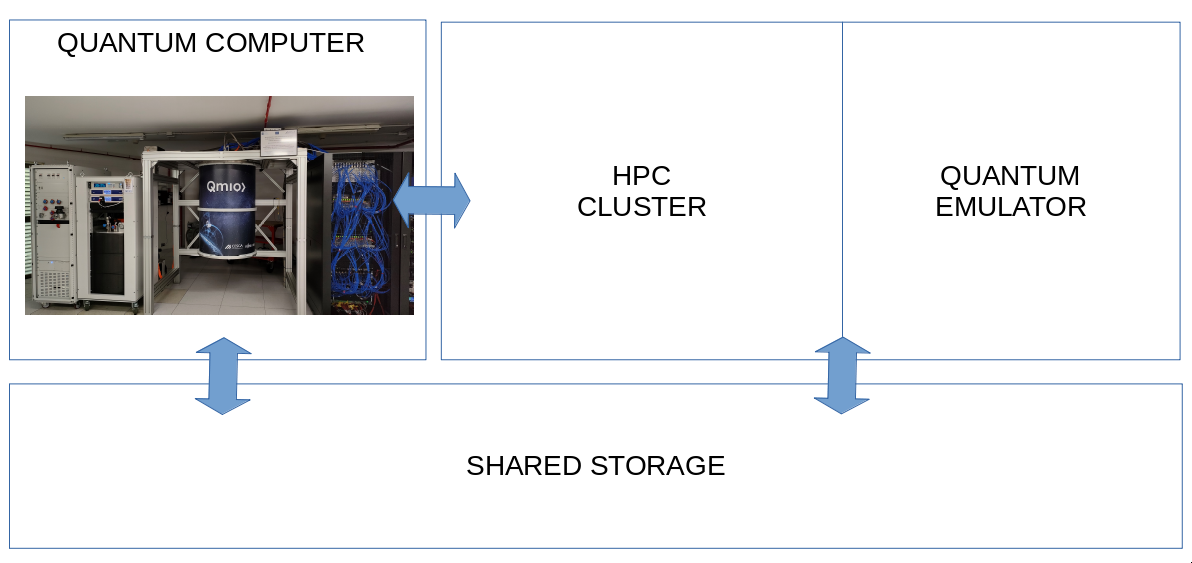}
    \caption{Overview of QMIO's hardware architecture. It is mainly composed of a HPC system, a quantum computer with a 32-qubit QPU, and classical quantum circuit emulator able to emulate up to 34 qubits. All the systems are locally interconnected and have access to a shared storage.}
    \label{fig:hardware}
\end{figure}

 QMIO is composed both from HPC and QC hardware co-located in the same data center -- a standard HPC data center -- which required some changes to accommodate this new hybrid computing infrastructure. 
\Fig{fig:hardware} depicts a schematic of the different hardware components involved in the HPCQC system.


\subsection{Data Center Facility}\label{hardware:facility}
Quantum computers are much more sensitive to noise than classical computers, resulting in strict requirements for controlling unwanted noise originating from various environmental factors.
A fundamental part for the correct functioning of a quantum computer is to keep the QPU as isolated from the outside world as possible. 


HPC data centers have advanced environmental control mechanisms in place to control the temperature and relative humidity\cite{mcfarlane2021ashrae}. 
Moreover, the electrical facilities include Uninterruptible Power Supply (UPS) systems that ensure electrical stability during grid changes, while also filtering out various electrical abnormalities.
As consequence, an HPC data center satisfies most of the requirements for the installation of quantum computing infrastructure because it provides a controlled environment, particularly with respect to temperature, humidity, and power supplies.
However, certain aspects were beyond these capabilities and required the application of several enhancements. 


In the initial state of QMIO deployment, all these aspects were taken into account. 
To ensure the stability of the quantum units, three main aspects were analysed: the physical location of the cryostat, signal control systems, and cryogenic equipment; the electrical power supply; and the room temperature control.

Regarding the placement of the necessary equipment, the building structure was verified and the floor was leveled off with an extra layer to better receive the heavy cryostat structure and to install the noise-isolation frame.
The electrical power supply was designed to fit the requirements of all pieces of the system (power supply redundancy, uninterrupted power supply using UPS systems and isolate compressor engines from the UPS-secured circuits), using UPS connections when needed and checking the quality of ground impedances to ensure the needs of the system. 
The environmental measurements of the room were taken into account and have been monitored live since then, permitting the research into correlations between system fidelities and room environmental conditions,
as suggested by other works \cite{kono2024mechanically}.

The QC system also required the establishment of a weekly supply of liquid Nitrogen to the data center\footnote{To decrease the temperature, the cryostat uses three components: Nitrogen, Helium-4 and Helium-3. Helium is used inside a closed circuit, but Nitrogen needs continuous replacement.}, this is required by the cryostat and stored in an open cage outside of the building.

\begin{figure}[tbh!]
    \centering
    \includegraphics[width=0.85\linewidth]{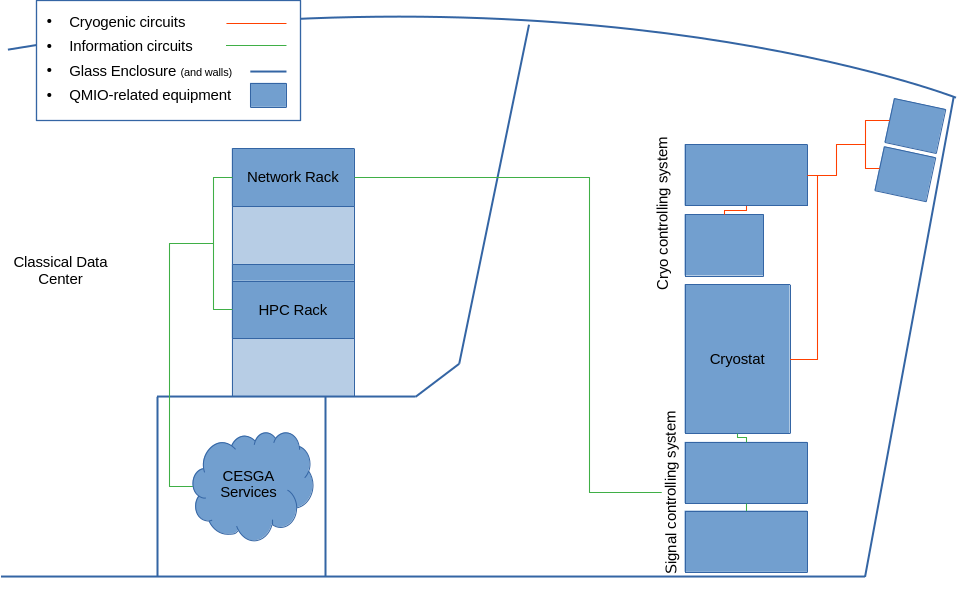}
    \caption{Diagram showing the location of the various components in the data floor. On the left side, the classic HPC components are located together with the quantum emulation machines and the storage and connectivity equipment. This equipment shares space with other CESGA clusters. On the right side, there are all the components that are responsible for ensuring the functionality of the quantum hardware. Both sides are divided by a glass wall and interconnected under the raised technical floor.}
    \label{fig:plano_qmio}
\end{figure}

All cabling was done under the raised technical floor using the existing paths of the data center. 
In \Fig{fig:plano_qmio}, you can see a scheme of the final layout and connections. 
As seen in the figure, a separation glass wall was also built to separate loud noise sources from the HPC part of the cluster and other CESGA systems from the quantum hardware infrastructure to minimize noise possible impact.

\subsection{HPC Cluster}\label{hardware:hpc}


Viewed from an overall perspective, QMIO employs a conventional High-Performance Computing cluster structure, including a set of \emph{admin nodes}, responsible for administrative functions, and \emph{login nodes}, which provide user access to the cluster. 

The system includes compute nodes of different architectures, managed separately at the orchestrator level. 
On one side, a partition consisting of 23 \emph{x86\_64} dual socket nodes, each with a total of 64 cores and 1 TB of main memory. 
This partition is called \emph{ilk}, with one node specifically dedicated to executing hybrid quantum-classical programs, as explained later. These nodes also provide interactive sessions to the users. 

On the other side, there is a set of 16 \emph{ARM} nodes, specifically Fujitsu's A64FX with 48 cores and 32 GB of HBM2 memory. This partition is called \emph{a64} and it is used mainly for quantum circuit emulation purposes. All nodes on both partitions are interconnected using a low-latency Infiniband network.

Finally, the system integrates a quantum computer based on superconducting qubits developed by Oxford Quantum Circuits (OQC), which will be discussed in depth in the following sections. A summary of all of these hardware components is presented in Table \ref{tab:qmio-cluster}.

\subsection{Quantum Computer}\label{hardware:qpu}

\begin{figure}
    \includegraphics[width=\linewidth]{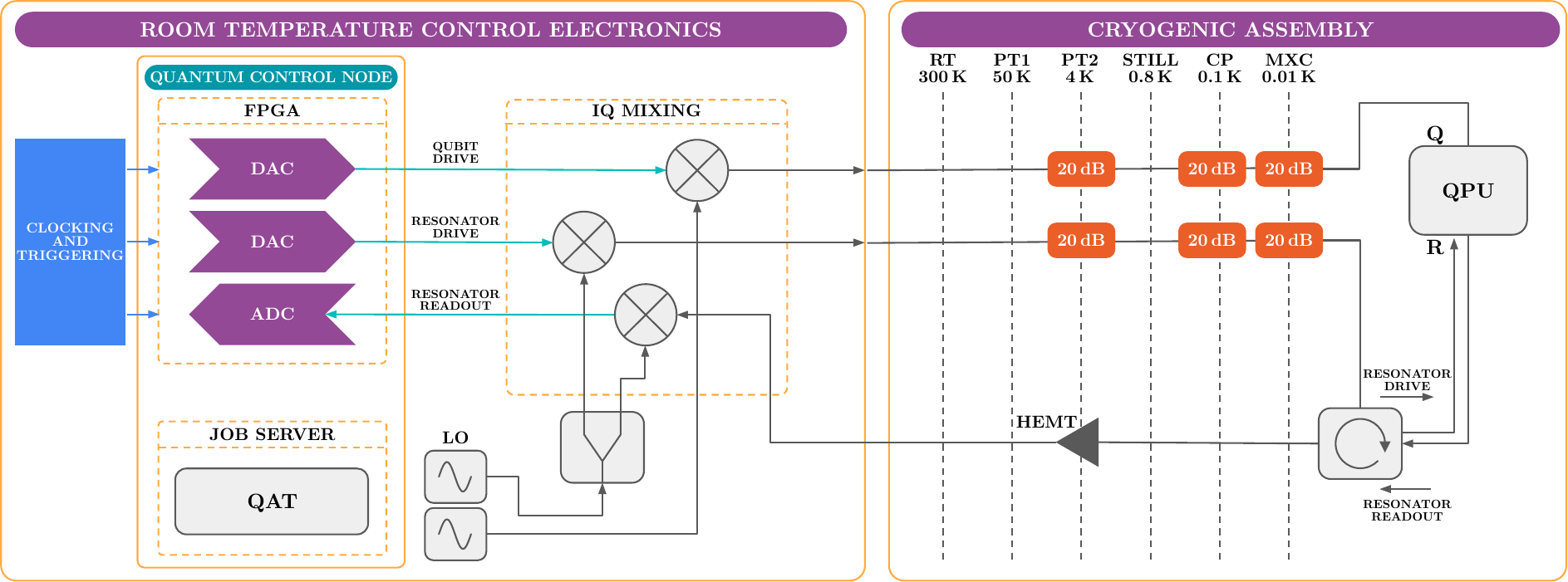}
    \caption{\label{fig:hardware-stack-summary} Schematic showing Left: Room-temperature control hardware as described in section~\ref{sec:ctrl-hw}; Right: Cryogenic assembly as described in section~\ref{sec:cryo}.}
\end{figure}

The general layout of the quantum hardware stack can be found in \Fig{fig:hardware-stack-summary}.
The Quantum Processor Unit (QPU) is operated in a milli-Kelvin environment, provided by a dilution refrigerator. 
Room temperature control electronics are responsible for the generation of qubit control and resonator readout signals and the measurement of qubit states. 
The cryogenic assembly provides interconnections and signal conditioning between the control electronics and the QPU housed in the dilution refrigerator.

\subsubsection{Quantum Processing Unit}\label{sec:qpu}

The QPU is based on the superconducting coaxmon architecture~\cite{coaxmon_paper}, consisting of 32 coaxmon unit cells patterned on sapphire in a lattice. 
To reduce interactions with electromagnetic modes, inherent to the enclosure housing the QPU, pillars are machined in the enclosure body~\cite{coaxmon_pillars_paper}. 
The pillars inductively shunt the enclosure, increasing the frequency of these electromagnetic modes to limit detrimental effects, including radiative energy relaxation and mediation of unwanted crosstalk within the enclosure~\cite{Spring2022}.
For more details how these pillars were added and the benefits, see \Refer{coaxmon_pillars_paper}.

The coaxmon qubits are OQC's patented design allowing out-of-plane driving and measurement~\cite{coaxmon_paper}.
All of the coaxmons are fixed-frequency transmons, where the frequency ranges between 4--6~GHz and anharmonicity of around 180--190~MHz.

The qubits are on top of the substrate and the resonators sit underneath, allowing for capacitive coupling between the two for measurement.
The qubits are driven and measured by pins that are placed out of plane, coming from the top and bottom respectively.
By having adjustable pins, the distance from the qubits can be optimized because it directly affects the external quality factor~\cite{coaxmon_paper}.


With all these connections off the chip, this leaves space to implement couplings between qubits, which are coupled capacitively.
Coupling arms come from each of the qubits and meet in between, with a small gap to allow capacitive coupling.

\subsubsection{Control of qubits}\label{sec:control}Control of the qubits starts from the Quantum Control Node (QCN), a classical computer where the quantum program instructions are converted to analog pulses via a Digital-to-Analog Converter (DAC) and mixed.

On measurement, a similar process occurs for the readout analog pulse (resonator drive), which goes through all the same stages, but using a different set of cables.
When reaching the QPU, the measurement pulse is applied to the resonator.
On measurement, a signal is sent back from the resonator and returns up the cryo-assembly, via a circulator that takes the signal down a separate resonator readout line (resonator readout).
This signal is then sent through the IQ mixing until ultimately converted back to a digital pulse via an Analog-to-Digital Converter (ADC) and returned for post processing.

\subsubsection{Cryo Hardware}\label{sec:cryo}

To avoid errors attributed to thermal excitation, the QPU package is housed in a milli-Kelvin environment, provided by a \textit{Oxford instruments Proetox LX} dilution refrigerator.
Interconnections between the package and room temperature control electronics are provided by a carefully constructed cryogenic infrastructure, inclusive of signal conditioning. 

Microwave drive signals, for qubit operations and resonator readout, are transmitted via highly-attenuated, low-thermal-conductivity, stainless steel, semi-rigid coaxial cables. 
Attenuation is distributed over several temperature stages of the platform, using thermally anchored cryogenic coaxial attenuators. 
The attenuators serve two purposes. 
Firstly, they allow for thermalization of the coaxial cables, constraining the passive heat loading on temperature stages. 
Secondly, they attenuate black-body radiation propagating via the cabling from higher temperature stages, ensuring the thermal photon occupation number at the qubit frequency is of order 10$^{-3}$ at the milli-Kelvin stage of the refrigerator, isolating the quantum processor from incident thermal noise to avoid additional sources of dephasing and energy relaxation~\cite{Krinner2029}. 

Prior to connection to the QPU, each cable arriving at the milli-Kelvin stage is interrupted with a combined low-pass and infrared non-magnetic filter. 
This removes any unwanted frequency components in the drive signal, with the potential to cause dephasing and blocks high-energy radiation reaching the QPU, which can result in cooper pair breaking and thus quasiparticle poisoning resulting in energy relaxation. 

The resonator drive signal, reflected from an individual qubit readout resonator, is routed via a cryogenic microwave circulator to an amplified output channel. 
The circulator is a non-reciprocal microwave component that provides reverse isolation, avoiding back-action on the QPU from the output channel. 
Low insertion loss superconducting Niobium Titanium semi-rigid coaxial cable is used to transmit the reflected resonator readout signal. 
This is placed between the milli-Kelvin stage and a higher temperature stage (4 K). 
The signal is then amplified by a low-noise high-electron mobility transistor (HEMT) cryogenic amplifier to increase the measurement efficiency, prior to being transmitted to room temperature and the associated electronics by a copper nickel coaxial cable.

\subsubsection{Control Hardware}\label{sec:ctrl-hw}
Control and measurement of a quantum computer requires electronic hardware capable of acting upon requested compute tasks by synthesizing Radio-Frequency (RF) pulse sequences, which perform the gate and readout operations on the QPU. 
High-fidelity control requires sub-nanosecond timing accuracy, synchronization between control channels, stable frequency generation, and low-noise signal generators~\cite{qpu_interface,clockstability}. To achieve this, QMIO uses OQC's control hardware solution. 

The QPU requires signals between 4--6~GHz for qubit operations and 8--12~GHz for readout, with phase and amplitude control for pulse-shaping~\cite{qe_bible}. 
This is enabled by using an industry-standard method: IQ mixers to up-convert pulsed sub-1~GHz Intermediate-Frequency (IF) control signals, generated by the DAC, to the required frequency band by mixing with a stable clock, known as the Local Oscillaror (LO). 
A similar setup is used for readout, down-converting the Radio Frequency (RF) signal to an IF band for digitization with the ADC. 
Unwanted spectral content is suppressed by introducing amplitude and phase imbalances to the IF output, and a DC bias to the IF ports suppresses LO leakage.

Each qubit uses a dedicated Field-Programmable Gate Array (FPGA) attached to the control server to handle generation of IF pulses using DACs for qubit and resonator control, and to receive the digitized readout signal from the ADC. 
A clocking system distributes a 2~GHz differential clock, with a single-ended trigger, used to align clocks between FPGAs and synchronize playback.

\subsubsection{Calibration}\label{sec:calibrations}
To ensure consistent performance of QMIO, automated calibration scripts are run daily on weekdays.
Calibration scripts are suspended over the weekend period to allow for longer duration runs, free of interruption and the analysis of system stability over a longer time constant.
See Table \ref{tab:calibrations} for a summary of calibration.

The standard daily calibration script calibrates the basics of qubit characterization, such as frequency, as well as the direct recalibration of the single- and two-qubit gates, such as the amplitude of the pulses.

These routines take around two hours to complete daily.
Automated reporting flags any parameters that do not meet a specified threshold. 
In this instance a 1-hour time period is reserved for OQC quantum engineers to manually optimize the calibration if needed.

Since it is important to allow much access time as possible, two longer scripts are performed just once a week.
These are used to calibrate the mixers and benchmark the system.
The benchmarking script runs through all qubits and qubit pairs to calculate the gate and readout fidelities.

\begin{table}
    \centering
    \begin{tabular}{|c|c|}
    \hline
        \textbf{Measurement} & \textbf{Timing}  \\
        \hline
        1- and 2-Qubit Calibration & Mon-Fri \\
        \hline
        $T_1$ & Mon-Fri \\
        \hline
        $T_2$ & Mon\\
        \hline
        Randomized Benchmarking & Mon\\
        \hline
        Mixer Calibration & Mon\\
        \hline
    \end{tabular}
    \vspace{.4cm}
    \caption{\label{tab:calibrations} Measurements taken each week. Daily, we perform full calibration of the qubits and the 1- and 2-qubit gates. We also measure the $T_1$ daily to measure the long-term stability. Every Monday, we do a full benchmarking of the system, this includes also measuring the $T_2$ of each qubit along with the fidelity via randomized benchmarking of all 1- and 2-qubit gates. Further, we perform a full calibration of the mixer every Monday.}
\end{table}


\subsection{Quantum Emulator}\label{hardware:emulator}
The quantum emulator system consists of two specialized and dedicated nodes that operate on a classical circuit emulation paradigm with distributed memory architecture. 
You can find a summary of this hardware in Table~\ref{tab:qmio-cluster}.

These two chassis, known as FX700, have 8 nodes, each equipped with an A64FX processor from Fujitsu. 
These processors feature a 48-core, 64-bit ARM architecture and support extended vector instructions and High-Bandwidth Memory (HBM2) on the chip, which is essential for achieving a performative quantum circuit emulation.

In this hardware, a distributed version of Qulacs (MPI-Qulacs~\cite{Suzuki_2021}) is deployed. This the open source quantum emulation software   is considered one of the most efficient existing quantum emulation frameworks~\cite{10313736}. This hardware-software solution permits the emulation of quantum circuits with up to 34 qubits. As consequence, it enables seamless integration of the emulator with heterogeneous workloads with the quantum hardware.

\begin{table}[h!]
\centering
\begin{tabular}{ll}
\hline
\hline
x86 Nodes & Primergy\\
\hline
\# of chassis & 23\\
\# of nodes & 23\\
Interconnetion & IB HDR\\
\# of cores & 1472\\
memory per node & 1TB \\
\hline
\hline
Emulator & FX700\\
\hline
\# of chassis & 2\\
\# of nodes & 16\\
Interconnetion & IB HDR\\
\# of cores & 768\\
memory per node & 32GB \\
max qubit count & 34\\
\hline
\hline
Quantum Hardware & Qmio\\
\hline
Qubit technology & Superconducting (Coaxmon)\\
\# of Qubits & 32\\
Topology & Hexagonal Latice - tripartite max\\
\# of channels & 3x[qubit-resonator]\\
\hline
\end{tabular}
\caption{QMIO hybrid cluster hardware summary.}
\label{tab:qmio-cluster}
\end{table}

\subsection{Shared Storage}\label{hardware:storage}
 
QMIO hardware design includes a shared storage filesystem between all the computing components of the infrastructure. Currently, it is provided by a dedicated Network Attached Storage (NAS) system via the Network File System (NFS). This is a way to efficiently exchange information between them when needed. In the future, the control server will be also connected to the  Infiniband network, which will permit to mount the HPC Lustre filesystem, providing a high performance shared storage.



\subsection{Connectivity}\label{hardware:connectivity}
The classical computers are interconnected with different networks.  The HPC and the quantum emulator nodes are interconnected with a low-latency InfiniBand network as well as Ethernet. The control server and the storage are only interconnected by Ethernet. The current configuration of the Ethernet network allows only one classical HPC node to connect to the control server. This configuration is due to the software stack configuration described in the next section.


\section{HPCQC Software}\label{sec:software}
\begin{figure}[tbh!]
    \centering
    \includegraphics[width=0.85\linewidth]{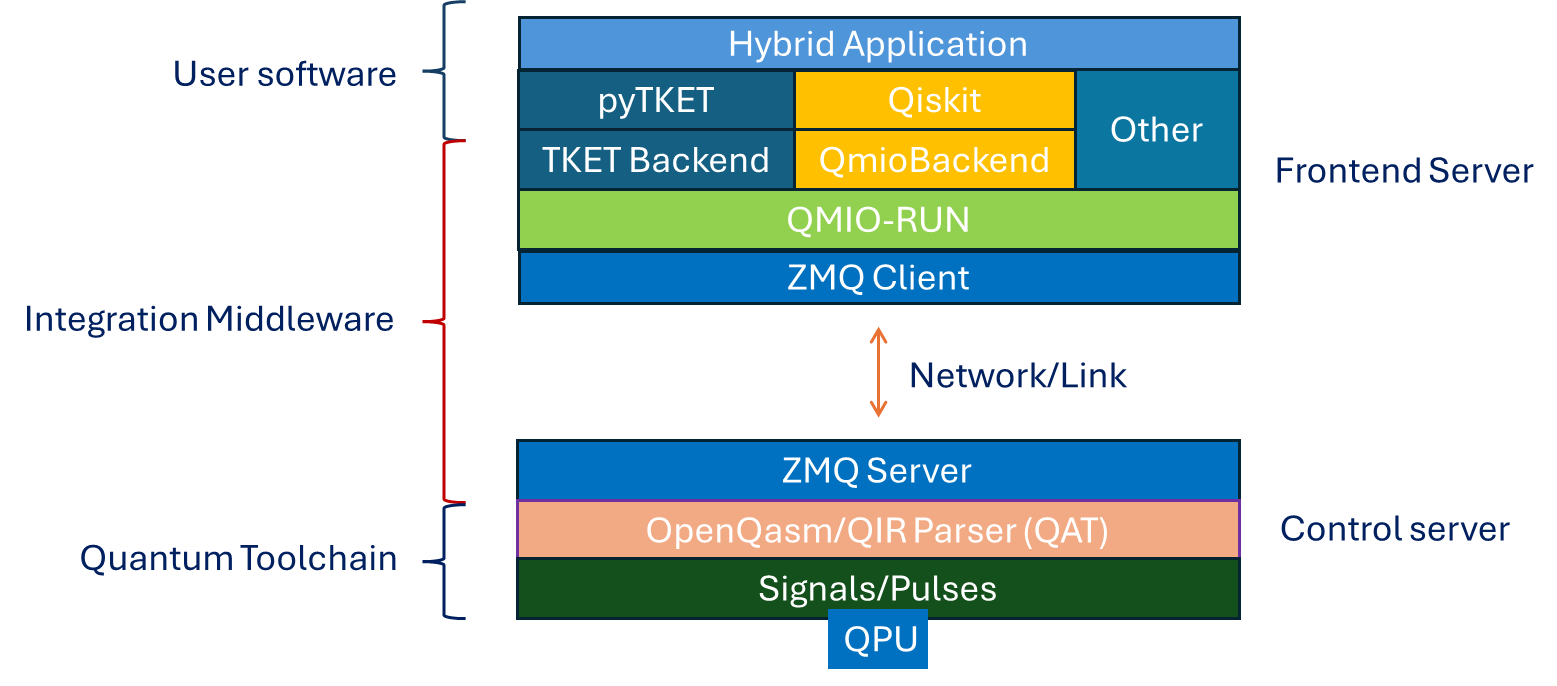}
    \caption{HPCQC software stack overview. On the top of the stack are the user applications written in high level frameworks like Qiskit. The integration middleware abstracts away the details of the underlying infrastructure. On the bottom layer the quantum toolchain takes care of all the low-level operations.}
    \label{fig:software_stack}
\end{figure}

The integration of HPC and QC offers new computational capabilities that require a dedicated software stack, see \Fig{fig:software_stack}.

At the top of the stack, the end-user software.
Below it, the integration middleware abstracts away the low-level details of the quantum system. It has been developed to allow for a seamlessly integration between the classical and quantum parts of the system. 
Finally, at the underlying layer, the quantum toolchain that takes care of all the low-level operations needed for the execution in the QPU. These layers are described in the next subsections.

\begin{figure}[tbh!]
    \centering
    \includegraphics[width=1\linewidth]{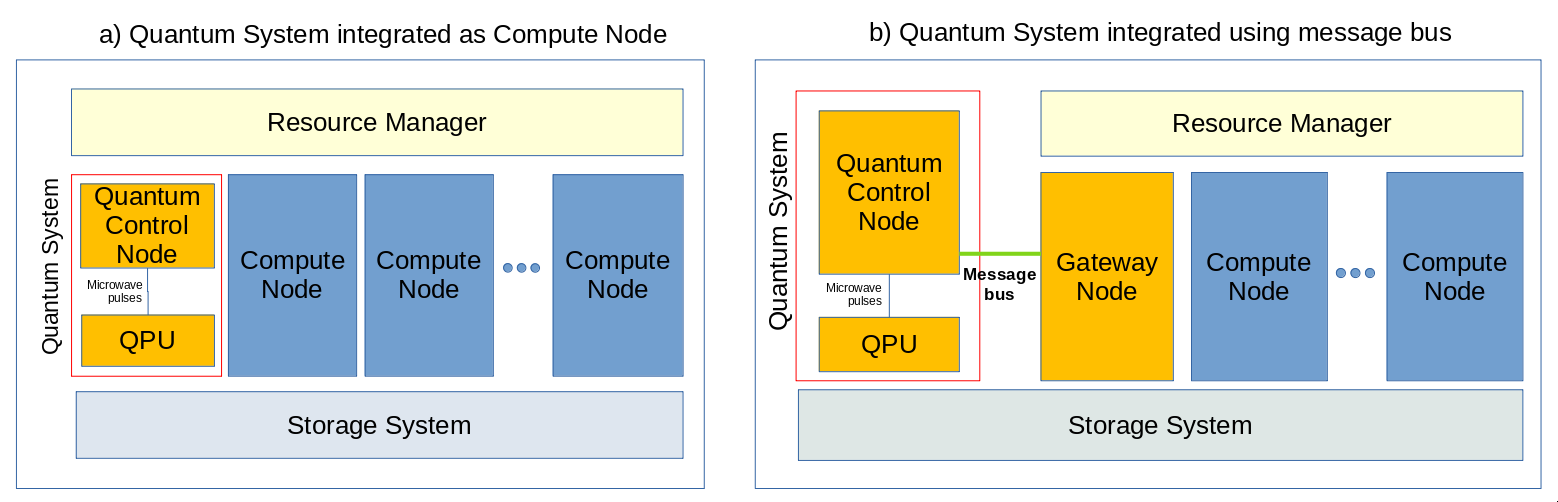}
\caption{HPCQC integration middleware design:  a) Integration of the Quantum Control Node as an HPC Compute Node. The Quantum Control Node is directly managed by the resource manager (SLURM). The Quantum Control Node runs a resource manager daemon (\emph{slurmd}) as any other HPC Compute Node. The resource manager allocates this node to the jobs that request to use a QPU. b) Quantum Control Node integration using a message bus: The Gateway Node is just a standard HPC Compute Node. The allocation of the Gateway Node is managed by the resource manager (SLURM). The Gateway Node is connected with the Quantum Control Node using a message bus, in our current implementation based on ZeroMQ.}
    \label{fig:qcp-as-hpc}
\end{figure}

\subsection{User Applications}\label{software:applications}
The user's software is deployed in the classical cluster following a conventional HPC approach using well established tools Lmod and Easybuild \cite{lmod_easybuild}.

This way the available software is offered to the users as Lmod modules that, once loaded, provide the correct environment for the application resolving the dependencies with other required modules. Also, this deployment permits the easy integration of several versions of the same software, allowing the users to execute the most convenient one for their programs.

Easybuild is used to compile the applications using recipes that allow to automate the process of re-compiling when a new version is available or when modifications are done to a given application.

The most common development frameworks for quantum computing are provided like Qiskit, PyTket, Qulacs or Pennylane as well as other classical software commonly used both in quantum computing and hybrid environments like Jupyter and different compilers and optimized mathematical libraries~\cite{uf_survey}. In fact, compilers are the critical tools, because the quantum programs (quantum circuits) must be optimized and transformed to use the QPU architecture and basic supported gates before executing.

\subsection{Integration Middleware}\label{software:middleware}
The integration middleware provides the underlying abstraction that user applications use \emph{under-the-hood} to transparently communicate with the QPU. So, its design and development has been a crucial part of the HPCQC system.

Nowadays integrating QPU and HPC systems is a challenge because they live in quite different worlds. Current HPC systems are clusters of a large number of computers forming a large single system with thousands or even millions of CPUs, sometimes improved with accelerators as the Graphical Processing Units (GPU), interconnected by low-latency networks like Infiniband. Commonly, they use a resource manager like SLURM to allocate available resources between the different users. Each user can submit several jobs, which can be executed concurrently but each one using exclusive resources (cores, memory and accelerators). Each job executes one or several user applications, which can expand several computers using  libraries like MPI to run in parallel. 


On the other hand, current quantum computers are systems that have just one QPU that is used to run quantum programs (usually known as circuits) generated by the user's application. In this case the common approach is to run each circuit independently even if there are enough free qubits to run another circuit at the same time. There is the possibility to pack several circuits but this can lead to unexpected behaviours due to factors like crosstalk. However, being this the current state-of-the-art, there are proposals for having more than one QPU per quantum computer (know as modular computer) or to have several QPUs, each one attached to a different classical node and linked by quantum networks, forming a High-Performance Quantum Computer and making Distributed Quantum Computing a reality~\cite{barral2024reviewdistributedquantumcomputing,Main2025}.


Nowadays, computers centers that deploy quantum computers have to cope with very different workloads. On one hand, there are some algorithms as QAOA, VQE or, in general, training of Parametric Quantum Circuits (PQC) as used by Quantum Machine Learning (QML) paradigm that demand frequent and fast execution of many circuits. These are long hybrid programs, where the execution of each circuit is managed by a classical algorithm that decides which are the parameters to use for their next iteration. For these algorithms, the circuits can be compiled advanced-ahead of time, only once, and executed many (even thousands) times. This characteristic demands that this compilation and optimisation of the circuits must be executed before the main workflow starts, in the same job or, preferable, in a separated previous job that does not have the QPU allocated. 

On the other hand, there are other users that want to execute small circuits to test new algorithms and ideas, that execute one o few circuits that demand small amount of time of the QPU.  
In the case of a quantum computer, each circuit execution with the current supercomputing qubit technology can last less than 500$\mu$s because after this time coherence is lost. However, in most cases the system will repeat its execution several thousand times to produce some statistics. Each execution is referred as one shot. So the execution time can be up to few seconds.

Taking these design constraints into account the initial design involved the direct integration of the Quantum Control Node as an HPC node. Even if that design \emph{a priori} looked as the best alternative, after being used several months, it did not meet user expectations due to the high latencies imposed in the job submission by the resource manager. This led us to re-design the integration middleware so that the Quantum Control Node is accessed through a Gateway Node using a message bus instead. The message bus demonstrated a much lower latency for task submission than the resource manager. 
\subsubsection{Integrating the Quantum Control Node as an HPC node}

The tightest possible integration of the Quantum Control Node (QCN) in the HPCQC system is to convert the QCN in another resource equivalent to any other HPC Compute Node. From the resource manager perspective, the QCN is just a node with a special accelerator that is the QPU, and it will schedule it as any other resource in the system. From the user perspective, the QPU is just a special resource that can be requested in a job, in a similar way as a node with a GPU would be treated.

The implementation involved installing the resource manager daemon---in our case \emph{slurmd}, the SLURM resource manager daemon---in the QCN and making all the user software stack described in Section \ref{software:applications} accessible from the QCN.

This way, when a user submits a job that requires the QPU, the resource manager allocates the QCN to the job. The job is then executed in the QCN allowing the user's application direct access to the QPU. 


During the time the user's application is running no other job can use the QPU. Since the application can run classical or quantum tasks, it is important to do as much of the classical tasks as possible before submitting the job. For example, one common classical task is the compilation of the quantum circuits to be run described in Section \ref{sec:qat}. If this compilation is done inside the job, then while the compilation is performed the QPU sits idle. So the middleware provided tools that allowed to compile circuits in advance---ahead of time compilation (AoT)---and we directed users to do that whenever possible.

To simplify usage we also implemented backends for common QC frameworks like Qiskit and PyTket that abstracted away the underlying middleware implementation. The backend took care of running the quantum circuits and returning the results to the user's application.

The main issue with this approach was that, each time the Qiskit or PyTket backend internally submitted a quantum circuit job to SLURM, it incurred a significant overhead---between 1 and 3 seconds---because the SLURM resource manager  is not intended to run jobs of less than a second of duration. This prompted us to explore alternative approaches.

\subsubsection{Integrating the Quantum Control Node using a message bus}

This second scenario makes a big difference for the resource manager, because HPC resource managers like SLURM are not expected to deal with such short computations and in fact, the scheduling process of a job can take several seconds. This makes quite inefficient to submit each circuit as a SLURM job and it would require a different type of resource manager than the one used in the HPC system. This scheduling latency overhead is also undesirable for the PQC executions. Due to the drift of the basic properties of the current superconducting QPUs, waiting times between the execution of two sets of circuits of a hybrid PQC algorithms could affect negatively the convergence of the algorithm. For this reason, PQC hybrid algorithms demand a continuous and exclusive access to the QPU during longer periods than single tests.

These two different scenarios can be coined as \emph{Batch usage}, when the user's program has a direct, usually unattended, and exclusive access to the QPU for long periods of time, and \emph{Interactive usage}, when the user demands short executions, usually from an interactive notebook while waiting for the result. 

Another important consideration is that a research infrastructure like QMIO must support a variety of user needs, from high-level programming frameworks as Qiskit or Pannylane, to low-level intermediate representations as QIR, or even pulse-level programming using OpenPulse. The software stack must be as agnostic as possible to these frameworks, especially in the layers closest to the QPU. This design choice will permit the evolution of the technology at the different layers as the current state-of-the-art advances. 

Finally, a last characteristic to take into account in the design of the interface between the HPC system and the QPU (or the Quantum Control Node) is the fact that they are basically isolated systems that communicate with each other through the network.




The QMIO integration middleware is designed to accommodate these different scenarios and requirements. Application frameworks such as Qiskit or pyTket are supported through specific interfaces (known as \emph{backends}), although end users can also develop their own interfaces or submit circuits written in the supported intermediate representations, currently QIR, and OpenQASM 2.0 and 3.0.

The middleware provides users with all the necessary tools to perform circuit transpilation and optimisation. The tools take into account qubit connectivity, the supported set of qubit gates, gate timings, gate fidelities, and coherence times (T1 and T2). This allows users to perform ahead-of-time (AoT) compilation, even making use of parametric circuits when needed (as is the case of QAOA, VQE or the iterative Quantum Phase Estimation). Both backends convert the proprietary circuit representations to supported open intermediate representations (QIR, and OpenQASM 2.0 and 3.0). 

Below this layer, a middleware component decouples the programming frameworks from the subsequent middleware layers. This component, known as \emph{qmio-run}, provides the \emph{QmioRuntineService}, which receives the generated intermediate representation along with additional execution parameters such as the number of shots, the desired output format, and the repetition period\footnote{Because QMIO does not support the reset to fiducial state of the qubits yet, between two shots the execution must be stopped to allow the QPU to relax all qubits to state $|0\rangle$. This waiting time has a default value but can be controlled by the user using the repetition period parameter, that is defined as the time that separates the first instruction of two consecutive executions of the program.}). It then transfers this data to the Quantum Control Node using a low-overhead communication layer based on ZeroMQ\footnote{https://zeromq.org/}. 

Once the message reaches the ZeroMQ broker in the Quantum Control Node, it is validated and passed to the QAT quantum toolchain for execution in the QPU (see section \ref{sec:qat}). After circuit execution and measurement, the results are sent back via ZeroMQ, received by the \emph{QmioRuntimeService}, and ultimately returned to the user-level framework.

\begin{figure}[h]
    \centering
    \includegraphics[width=0.85\linewidth]{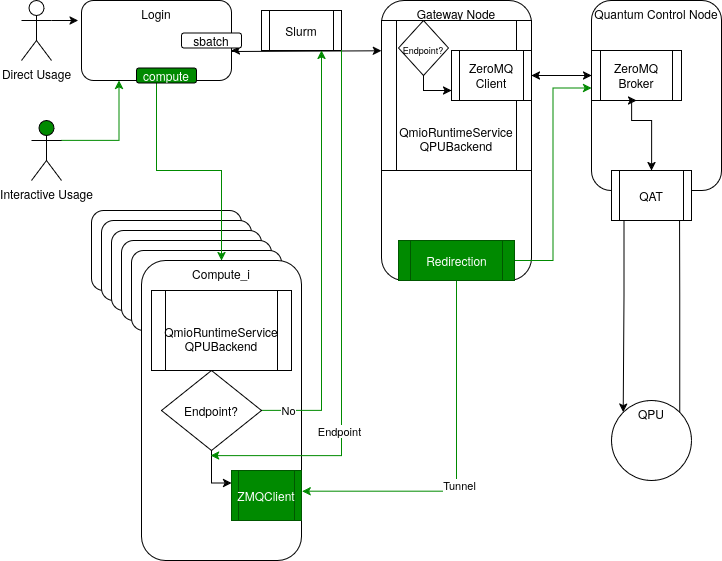}
    \caption{QMIO execution workflow diagrama. The user in \emph{white} asks for  hybrid resources in \emph{batch} mode (\emph{Batch Usage}).  The user in \emph{green} accesses them interactively (\emph{Interactive Usage}. Resource reservation and allocation is happening transparently to the end-user. A fast communication channel using ZeroMQ is opened and reused as many times as needed. When the resources are released, they become available for other jobs without disrupting any running tasks.}
    \label{fig:qmio_library}
\end{figure}

Using this software stack is possible to handle both direct and interactive use as shown in \Fig{fig:qmio_library}. On one hand, the Direct usage forces the user to ask SLURM for a job asking for the allocation of the QPU. This submission will allocate the resources on the HPC node (frontal node) with connection to the quantum computer, and the full unattended job will be executed. Currently, the maximum allocation time for a single job is two hours, although this limit can be changed in the future. The node is allocated in exclusive for this job to avoid any delays in the classical part of the algorithm. 

On the other hand, an Interactive process (green line in the figure) uses the job allocation as well. However, for interactive sessions, user should previously ask for resources through the command line and SLURM will allocate them into specific ilk nodes prepared to fulfill this purpose. These nodes are commonly shared among different users and do not have direct access to the QCN. This access will be handled exclusively through the frontal node as for the \emph{Direct usage}, but this request is made by the QMIO-Run that does this process completely transparent for the user. When the interactive session is allocated, a Jupyter notebook or some other interactive interpreter is spawned and the routine preparation starts. When the instructions that require results from the quantum system are reached, the QMIO-Run handles the communication with SLURM, to request a job in the frontal node to route the message pass-through. Obviously, this will add an overhead if some other user is running a job with the quantum resources. Once this new job starts, route through the frontal node is provided to transfer all the messages to/from the quantum computer. Once the routing is established a socket-socket communication goes on until finishing the connection or reaching a small time limit (the maximum allocation time for this method is constrained to few minutes). This improves user concurrency without the addition or big overheads. The use of the socket-socket communication over the frontal node route does not add any significant increase in runtime.


\subsection{Quantum toolchain}\label{sec:qat}
 The underlying layer of the HPCQC software stack is the quantum toolchain or quantum compiler. It takes care of generating the control hardware instructions from the logical program, executes the program and finally performs post-processing on the results before returning them to the user. Its role is not limited to just produce executable code as in the case of a classical compiler. 

QMIO control server uses the Quantum Assembly Toolchain (QAT), an open-source quantum compiler developed by OQC\footnote{Available at https://github.com/oqc-community/qat}. QAT accepts a quantum program in an intermediate representation (as said before, nowadays supports QIR, OpenQASM2.0 and 3.0), optionally does logical compilation via Tket (but QMIO users are advised to compile and optimise the quantum programs before off-loading them to the QPU, because this step could be extremelly demanding, and this feature could be disabled in the future), generates the control hardware instructions from the logical program, executes the program and finally performs post-processing on the results before returning them to the upper layer. 

There are three central elements to the architecture of QAT, hardware models, instruction builders and engines. Hardware models are data structures that contain all data about a particular QPU. This includes calibration values for each gate, qubit couplings and calibrated readout values. Instruction builders are responsible for construction quantum programs and experiments. The instruction builders allow for custom pulses, but their main responsibility is to convert a logical instruction into a lower-level representation of pulses. Finally, the engines manage control hardware instruction lowering and execution via the control hardware driver. The hardware model and instruction builder are both serialisable, this allows for pre-compilation to improve throughput. 

QAT provides another capabilities to be executed on QMIO normal nodes, as an echo hardware that always returns a default value which allows for testing of the compiler isolated from the real quantum hardware. There is also a real time chip (RTCS) simulator hardware model which allows for pulse level simulations. RTCS is an extension of QuTiP~\cite{Lambert2024} and allows for testing of OpenPulse programs and new calibration routines.


\begin{figure}[tbh!]
    \centering
    \includegraphics[width=0.95\linewidth]{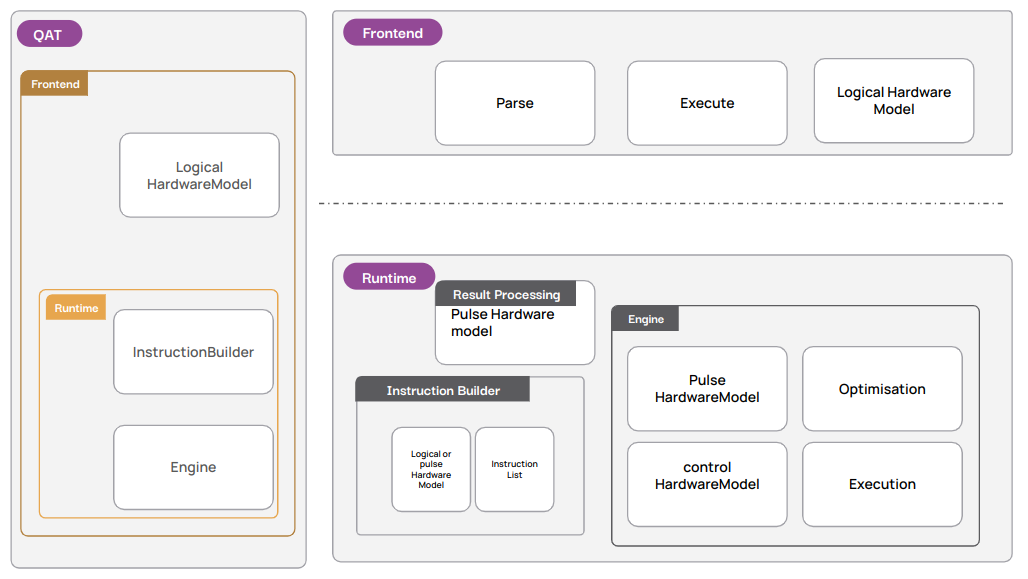}
    \caption{Architectural overview of the QAT compiler. The frontend is responsible for the parsing and optimisation of quantum programs. This processes generates and intermediate representation that is then passed to the runtime and backend which is responsible for lowering the intermediate representation to control hardware specific language, generating calibrated pulses and managing the runtime of the quantum program.}
    \label{fig:qat}
\end{figure}

\section{System Operation}\label{sec:operation}
Due to the intricacies of hardware and software infrastructure decisions, specific conditions emerge during daily and weekly operations, demanding careful attention.

Calibration is a crucial aspect of ensuring the QMIO system's optimal performance. 
Like other quantum systems, it cannot sustain good fidelity for indefinite periods. 
To mitigate this, the system undergoes daily calibration to maintain its performance standards, details of which are given in \Sec{sec:calibrations}. 
Additionally, a comprehensive weekly benchmarking process is performed to provide valuable insight into the system's evolution and improvement. 
This data is used at compilation time to inform decisions on algorithm placement and lower-level optimizations. 
The metrics are tracked daily to identify potential points of failure in the quantum hardware as soon as possible.

Monitoring is an integral component of this process, with a focus on gaining a deeper understanding of the system's capabilities.
Based on the calibration metrics provided by OQC, additional data is collected to place greater emphasis on specific system functionalities. 
These metrics are also supplemented by critical infrastructure data, including cryogenic system metrics and environmental variables such as room climate conditions and power supply readings.
This monitoring infrastructure was developed ad hoc to cater to the unique needs of this complex system, and its detailed explanation will be presented in a forthcoming dedicated work.

Leveraging our calibration and monitoring capabilities has revealed strong correlations between variations in system fidelity across different events. 
Notably, specific occurrences that mirror real-world implementation scenarios exhibit significant relationships with adjacent civil works, visits to the Data Floor, or other building interventions. 
These correlations require further investigation to fully understand their impact on overall system performance.

\begin{figure}[h]
    \centering
    \includegraphics[width=0.95\linewidth]{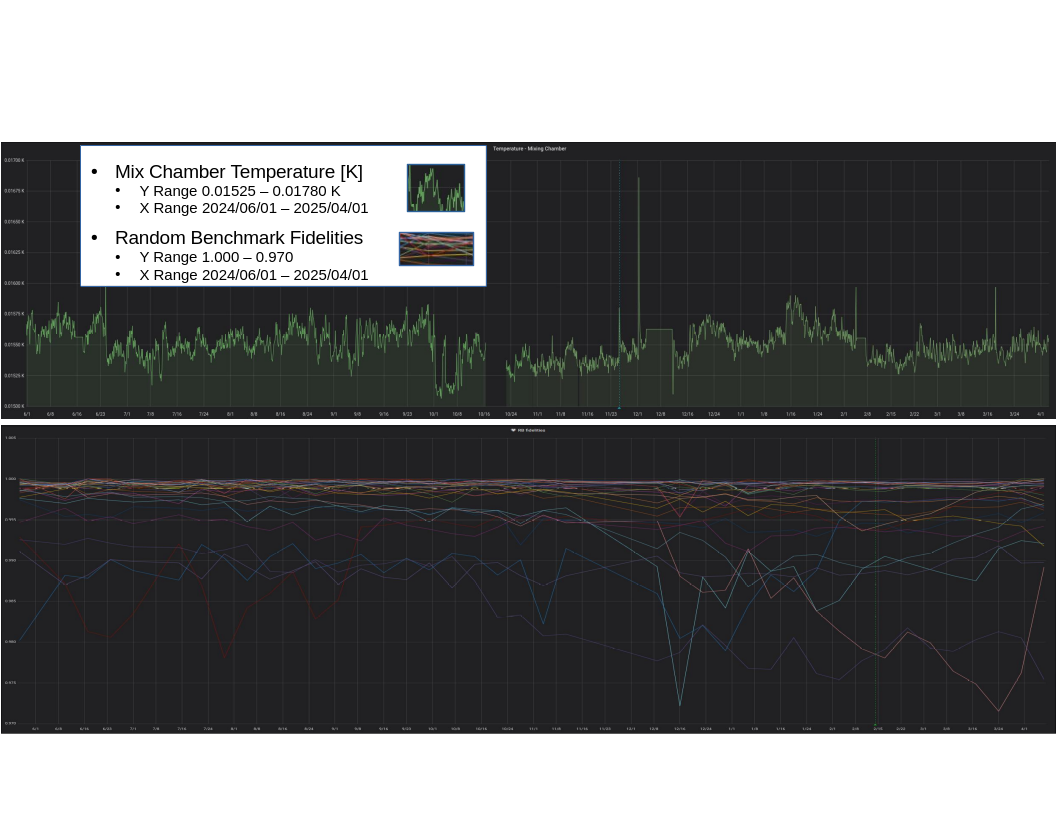}
    \caption{Example view of 10 month of monitoring data recovered from QMIO. Top part: Temperature in the mix chamber. Bottom part fidelities of randomized benchmarking}
    \label{fig:10months}
\end{figure}


\section{Discussion}\label{sec:discussion}
The development of quantum computers has been progressing rapidly during the last years and, although we are still in the NISQ era, it has already achieved important breakthroughs. 
Commercial quantum computers are already a reality and it is now possible not only to access them on-cloud but also to deploy them on-premises. 
This enables HPC centers to deploy quantum computers locally and start exploring how to integrate them with HPC systems. 
Several HPC centers are already deploying or planning to deploy QC technology, so we expect that there will be many new hybrid HPCQC systems like QMIO in the coming years.

QMIO started its operation in October 2023~\cite{qmio_inauguration} and it has been running in production, open to worldwide users, from May 2024~\cite{qmio_open,qmio_qcalls}. 
During this time we had the opportunity to design, implement and test different HPCQC integration approaches that were ultimately validated by the end-users of the system.

We started with a design that integrated the Quantum Control Node as an HPC node, allowing the quantum computer to be treated by the scheduler like any other HPC node in the cluster. 
This was the tightest integration possible; although it had several advantages, such as the direct execution of applications in the quantum control hardware, it presented an important disadvantage for the classical part of the code, which had far fewer resources available in the QCN than in the HPC node. 

In the second iteration we evolved the design, resulting in the addition of a dedicated HPC node that acts as a gateway for the QCN. 
This still enables us to use only one scheduler---the HPC cluster scheduler---and provides better isolation of the classical code that now has access to more computing resources---memory and compute---as well as to the low-latency network for MPI communication.

This does not mean that the first design is not valid, simply it did not work well in the current setup and for the type of applications run by the users (for example, the training of parametrized quantum circuits that need frequent access to the QPU). 
Its limitations could be overcome by redesigning the QCN node to make it look more like an HPC node---more computing power and memory, and low-latency network---and creating a better isolation of the classical software that would run there. 
To achieve these goals there is a need for a close collaboration between QC vendors and HPC centers.

From the user's perspective, the second design offers better performance and more flexibility. 
From the system's operation point of view, this design has allowed us to simplify the upgrading of the middleware, allowing it to be done in a transparent way for the users without requiring changes in their codes.

Another aspect to consider is the generality of the design presented. 
Even if our middleware was designed by focusing on the specific characteristics of QMIO---i.e. a superconducting QPU and no low-latency network in the QCN---we think that the middleware developed could accommodate other QC technologies as well as several QPUs. 
In the same way as it is done with the current QPU and QCN, each additional QCN will be interfaced through a dedicated HPC node, so they will be seen by the scheduler as an additional resource with support for QC that could be allocated as any other HPC resource.

\section{Future Work}\label{sec:further}
Looking ahead, several promising avenues remain for the continued development and refinement of our hybrid HPC-quantum computing infrastructure. One key area is inter-QPU communication, where we aim to enable seamless information transfer between quantum processing units (QPUs). Specifically, this would involve mechanisms for one QPU to pass measurement outcomes or intermediate results directly to another, potentially enabling more complex and distributed quantum algorithms.

Another important direction is the implementation of parametric compilation techniques, which will be particularly relevant for optimizing families of parameterized quantum circuits. This would significantly reduce compilation overhead and allow for more efficient execution of variational algorithms and real-time adaptive quantum computations.

Additionally, we are actively developing a quantum monitoring and benchmarking framework that will provide deep visibility into system behavior across both quantum and classical components. This framework will support comprehensive performance evaluation, enabling us to identify bottlenecks and fine-tune both the quantum and HPC layers of the system. Through exhaustive benchmarking and performance optimization, we aim to maximize the utility, reliability, and scalability of the hybrid infrastructure.

Going beyond the current capabilities of the quantum units, future programs will need millions of quantum instructions, needing a tight collaboration between the classical and quantum processors. For this reason, there are some initial proposals to use some programming paradigms as OpenCL~\cite{Vzquez-Prez2024}. 

These efforts collectively represent the next steps toward a robust and production-ready HPC-quantum ecosystem, capable of supporting increasingly complex and computationally demanding workloads.

We do not know which breakthroughs will take place in the next years or which technology will dominate QC in the future, but we hope that the work presented in this paper could give us a glimpse of how next generation HPCQC systems will be and what are the challenges ahead that will have to be solved.

\section*{Acknowledgments}
This research project was made possible through the access granted by the Galician Supercomputing Center (CESGA) to its QMIO quantum computing infrastructure with funding from the European Union, through the Operational Programme Galicia 2014-2020 of ERDF-REACT-EU, as part of the European Union’s response to the COVID-19 pandemic. A. Gómez was supported by MICINN through the European Union NextGenerationEU recovery plan (PRTR-C17.I1), and by the Galician Regional Government through the “Planes Complementarios de I+D+I con las Comunidades Autónomas” in Quantum Communication. We extend our thanks to the entire OQC team for their contributions to QMIO, which was instrumental in this work, to the CESGA and Fujitsu technicians that participated in the deployment and configuration of QMIO. Special thanks to Peter Leek, Joseph Bilella and Matthew Heath for their review of this manuscript.

\section*{OQC Authors}
Ailsa Keyser, Anirban Bose, Apoorva Hegde, Benjamin P Rogers, Bryn Bell, Darren Hayton, Jonathan Burnett, Kajsa Eriksson Rosenqvist, Owen Arnold, Philip Clarke, Richard Bounds, Ryan Wesley, Samuel Earnshaw, Travers Ward, and  William Howard

\section*{CESGA deployment team}
Juan Villasuso, Carlos Fernández Sánchez, José Carlos Mouriño Gallego, Natalia Costas, Pablo Rey, Jorge Fernández Fabeiro, Fernando Bouzas, María José Rodríguez Malmierca, and Ignacio López Cabido

\bibliographystyle{splncs04}
\bibliography{biblio}

\end{document}